\documentclass[a4paper,10pt]{article}
\usepackage[pdftex]{graphicx}
\PassOptionsToPackage{hyphens}{url}\usepackage{hyperref}
\usepackage{amsmath,amsfonts,amssymb}
\usepackage{fullpage}
\usepackage{authblk}
\usepackage{setspace}
\usepackage{subcaption}
\usepackage{booktabs}
\usepackage{tabularx}

\usepackage[explicit]{titlesec}
\usepackage{sidecap}
\usepackage{pbox}
\usepackage[superscript]{cite}

\usepackage{amsfonts}
\usepackage{amsmath}
\usepackage{multirow}
\usepackage{longtable}
\usepackage{makecell}
\usepackage{graphicx}
\usepackage{xcolor}
\usepackage{comment}
\usepackage[utf8]{inputenc}
\usepackage[russian,english]{babel}
\usepackage{CJKutf8}

 \usepackage[utf8]{inputenc} 
\DeclareUnicodeCharacter{200B}{{\hskip 0pt}}

\date{}
\title{\bf{The illicit trade of COVID-19 vaccines on the dark web}}

\author[1,+]{Alberto Bracci}
\author[1,2,+]{Matthieu Nadini}
\author[3]{Maxwell Aliapoulios}
\author[3]{Damon McCoy}
\author[4]{Ian Gray}
\author[5,6]{Alexander Teytelboym}
\author[7]{Angela Gallo}
\author[1,2,8,*]{Andrea Baronchelli}

\affil[1]{\small City, University of London, Department of Mathematics, London EC1V 0HB, UK}
\affil[2]{\small The Alan Turing Institute, British Library, 96 Euston Road, London NW12DB, UK}
\affil[3]{\small Center for Cybersecurity (CCS), New York Univ. Tandon School of Engineering, Brooklyn, NY 11201, USA}
\affil[4]{\small Global Intelligence Team, Flashpoint. New York, NY 10003, USA}
\affil[5]{\small Institute for New Economic Thinking, Oxford Martin School, University of Oxford, Oxford OX2 6ED, UK}
\affil[6]{\small Department of Economics, University of Oxford, Oxford OX1 3UQ, UK}
\affil[7]{\small Department of Finance, Cass Business School, London EC1Y 8TZ, UK }
\affil[8]{\small UCL Centre for Blockchain Technologies, University College London, UK}
\affil[+]{\small Contributed equally}
\affil[*]{\small Corresponding author: abaronchelli@turing.ac.uk}

\begin{document}
\maketitle
\begin{abstract}
Early analyses revealed that dark web marketplaces (DWMs) started offering COVID-19 related products (e.g., masks and COVID-19 tests) as soon as the COVID-19 pandemic started, when these goods were in shortage in the traditional economy. Here, we broaden the scope and depth of previous investigations by analysing 194 DWMs until July 2021, including the crucial period in which vaccines became available, and by considering the wider impact of the pandemic on DWMs. First, we focus on vaccines. We find 250 listings offering approved vaccines, like Pfizer/BioNTech and AstraZeneca, as well as vendors offering fabricated proofs of vaccination and COVID-19 passports. Second, we consider COVID-19 related products. We reveal that, as the regular economy has become able to satisfy the demand of these goods, DWMs have decreased their offer. Third, we analyse the profile of vendors of COVID-19 related products and vaccines. We find that most of them are specialized in a single type of listings and are willing to ship worldwide. Finally, we consider a broader set of listings mentioning COVID-19  as proxy for the general impact of the pandemic on these DWMs  . Among 10,330 such listings, we show that recreational drugs are the most affected among traditional DWMs product, with COVID-19 mentions steadily increasing since March 2020. We anticipate that our effort is of interest to researchers, practitioners, and law enforcement agencies focused on the study and safeguard of public health.
\end{abstract}

\section*{Introduction}

COVID-19 has caused a worldwide economic and public health crisis, that demanded and stimulated a global response. Hundreds of possible COVID-19 vaccines have been proposed~\cite{WHO_COVID_Vaccines} since the first officially approved vaccines in late 2020, like Sputnik~\cite{burki2020russian} and Pfizer/BioNTech~\cite{ledford2020uk, tanne2020covid, Roberts2020}.
The subsequent initial scarcity and unequal distribution of COVID-19 vaccines~\cite{vaccine_inequality} have generated concerns about illicit trade early on. Interpol warned about illicit offering of COVID-19 vaccines already on December 2, 2020~\cite{INTERPOL_COVID_Vaccines}, while Europol confirmed the sale of fake COVID-19 vaccines on dark web marketplaces (DWMs) on December 4, 2020~\cite{EUROPOL_COVID_Vaccines}, which ``may pose a significant risk to public health''.  *Understanding how DWMs reacted to the demand for vaccines is therefore crucial to allow policy and public health agencies to be prepared and effectively counteract these threats in the future
 
Interpol and Europol's concern were validated by early research showing that DWMs have been an important channel to access online illicit trade during the pandemic, with masks, COVID-19 tests, and alleged medicines consistently advertised on these platforms. In a first report~\cite{broadhurstavailability}, $222$ COVID-19 related unique listings were registered on April $3rd$, $2020$ in $20$ DWMs. In our previous work~\cite{bracci2020covid}, 788 COVID-19 related listings were observed $9,464$ times between January 1, 2020 and November 16, 2020 in $30$ DWM, showing how DWMs swiftly reacted to shortages and public attention by offering sought products like masks and hydroxychloroquine. More recent reports, carried by the Global Initiative and Europol, have suggested that the overall structure of illicit online trading has gained significant benefits from COVID-19~\cite{GIATOC, EuropolEMCDDA}.  

DWMs are an ideal venue to participate in online illicit activities. They can be easily accessed via specialized browsers, e.g., Tor~\cite{dingledine2004tor}, that hide they identity and location of their users, and offer a variety of illicit goods including drugs, firearms, credit cards, and fake IDs~\cite{GwernDarkNets}. DWMs drew the attention of hundreds of thousands buyers and sellers over the years, with a trading volume that rapidly reached hundreds millions United States dollars (USD) per year~\cite{world2019world, elbahrawy2019collective}. The growing popularity of DWMs has attracted the interest of the scientific community, security researchers, and law enforcement agencies. The scientific community has explored the behaviour of DWMs users through comparative analyses~\cite{barratt2014use, martin2014lost, aldridge2014not, dolliver2015criminogenic, dolliver2015evaluating, broseus2016studying,bracci2021macroscopic} and case studies~\cite{van2013silk, van2014responsible, lacson201621st}. Law enforcement agencies have successfully closed several DWMs, seizing millions of USD, and performing dozen of arrests~\cite{Operation_Onymous, FBIAlphabay, WallStreetMarket, SizedBerlusconiMarket, BillionFedsSilkRoad, JokerStashDisruption2020, SILKROADSEALED}. However, DWMs are intrinsically resilient to these interventions~\cite{elbahrawy2019collective},  also thanks to the emergence of decentralized trade around them~\cite{nadini2022emergence}, and 2020 has been a record year for their revenue~\cite{record_year1,record_year2}.

Here, we report on our analysis of 194 DWMs until July 22, 2021. 
Throughout the pandemic, we detected a total of 10,330 unique listings that were directly affected by COVID-19, i.e., mentioning COVID-19 either in their body or title. Among these listings, 250 were offering vaccines.
 It is important to note that a listing does not correspond to the sale of a unit, as sometimes happens for example on Ebay, but corresponds to the availability of multiple units of a product, similarly to what happens for example on Amazon.  
Listings related to approved vaccines were initially detected on the Invictus marketplace starting from November 17, 2020, almost 2 weeks before their official approval. Also, listings offering a fabricated proof of vaccination were registered on the Hydra marketplace since February 15, 2021. These listings replaced previously identified COVID-19 related products, like PPEs, COVID-19 tests, and guides on how to illicitly obtain COVID-19 relief funds. The availability of these products have decreased with respect to previous observations, with only 187 listings detected between November 2020 and July 2021 against the 788 registered between January and November 2020~\cite{bracci2020covid}. Many vendors selling these products are highly specialised in only a type of product and willing to ship worldwide, thereby increasing the number of potential customers. By analysing all listings mentioning COVID-19, we assess the overall impact of COVID-19 on DWMs. We show that drugs are the only traditional DWMs product to have been indirectly, and increasingly, affected by the pandemic,  with vendors mentioning both pandemic related supply issues and delays  .

Our results confirm the concerns that several international agencies have expressed regarding the online illicit trade of  COVID-19 vaccines, and corroborate the link between the shortage of and public attention on medical products, and their availability on DWMs  . In addition, they reveal that DWMs were only partially affected by the pandemic, with mostly drugs related listings explicitly mentioning COVID-19, while other traditional DWMs products, like firearm and fake IDs, were not. To reach a large audience beyond academia, we released a website~\cite{monitoring_effort}, where we are providing constant updates on the effect of the pandemic on DWMs.

\section*{Data and methods}
\label{sec:methods}

Our dataset includes the most popular DWMs in 2020 and 2021, such as White House, Empire, Hydra, and DarkMarket~\cite{darknetstats_live_markets, broadhurstavailability} and was gathered by Flashpoint~\cite{flashpoint}, a company specializing in online risk intelligence. Note that the landscape of active DWMs is constantly changing: Empire exit scammed on August 23, 2020~\cite{Empire_exit_scam}, while DarkMarket was shut down by Europol on January 12, 2021~\cite{Darkmarket_Shut_Down}. 
The dataset was obtained by web crawling DWMs, which consists of extracting and downloading data from these websites. To this end, the web crawling pipeline has to overcome strong CAPTCHAs~\cite{ball2019data} and authenticate into the DWMs of interest. Downloading content from DWMs remains a challenging task, and the objective becomes even harder when the research study requires monitoring multiple DWMs for an extended period of time. Previous research groups have tried establishing a web crawling pipeline through a combination of PHP, curl, and MySQL~\cite{baravalle2016mining}, through the Python library Scrapy~\cite{celestini2017tor}, and through an automated methodology using the AppleScript language~\cite{hayes2018framework}. Despite these efforts, only a few open-source tools are available~\cite{decary2013datacrypto, ball2019data} for crawling DWMs. Therefore researchers, companies, and federal agencies often rely on commercial software, like X-Byte~\cite{xbyte}, and specialized companies, like Flashpoint~\cite{flashpoint}, to crawl DWMs. 

\begin{table}[!t]
      \begin{tabular}{l}
        \hline
          (i) Vaccine related set of keywords \\ \hline
        antibod, vaccin, antidot, vacun, immun, \begin{otherlanguage*}{russian} Инокул, вакцин, прививк, Ревакцин, Инокул\end{otherlanguage*}, \\
        \begin{CJK*}{UTF8}{gbsn} 疫苗, 反\end{CJK*}, impfstoff, Gegenmittel
​         \\ \\ \hline
        (ii) COVID-19 and brands related set of keywords  \\ \hline
        covid, corona,\begin{otherlanguage*}{russian} ковид, Коронавирус, Пандеми, Вирус, Спутник V, Инфекци, Симптом\end{otherlanguage*}, \\
        \begin{CJK*}{UTF8}{gbsn} 新冠病毒, 武汉肺炎\end{CJK*}, couronne, pfizer, astrazeneca, moderna \\
         \hline
      \end{tabular}
\caption{\textbf{Search of COVID-19 vaccines.} Keywords used to pre-select vaccine listings from the original dataset. Words are truncated to include different suffixes (e.g., vaccin yields vaccine, vaccination, vaccinate, etc.)}
\label{keywords}
\vspace{1cm}
\end{table}

Our DWMs dataset is used to complement and extend the analysis we previously performed for the period between January 2020 and November 2020~\cite{bracci2020covid}. We increase the analysed period by adding several more months, until July 22, 2020, and observe the evolution of COVID-19 related products over the second part of the pandemic. We also add several new DWMs, increasing their number from 30 to 194, and comprehending a total of 10.8 million unique listing titles. Only 84 of these DWMs mentioned COVID-19, 20 DWMs offered COVID-19 related products, and 19 vaccines, see Table~\ref{market_details}. Each unique listing is observed at most once per day. 

During the  considered period of   the COVID-19 pandemic, the illicit offer of vaccines constitute d one of   the biggest threats for global public health. We therefore use a method to detect vaccine that ensures the highest possible coverage and accuracy. From the listings, we considered two different text fields: the title and the body (that is, the listing's detailed description). We then pre-selected all listings which contained, either in the body or title, at least one word from two different lists of keyword. These lists of keywords are shown in Table~\ref{keywords}: the first list contains keywords related to vaccines; the second list contains keywords related to COVID-19 or vaccine brands like Pfizer/BioNTech. Note that using keywords like ``antibod'' or ``vaccin'' allows to match all words including these sets of strings, such as, antibode, antibodes, vaccine, vaccines, vaccinations, and so on. We considered several different languages, such as, English, Russian, Chinese, and German. Afterwards, we manually inspected the listings to exclude false positives from the dataset, we categorized the listings in specific subcategories (e.g. specific brands), and we standardised the analysed attributes for the analysis. For example, we converted all prices to USD at the daily exchange rate at the time of observation.

\begin{table}[]
    \centering
    \begin{tabular}{ll}
\hline 
Category & Keywords \\
\hline
Guides on scamming   & guide, fraud, exploit, scam, loan, relief, scampage, cashout  \\
Medicines            & chloroquin, azithromycin, favipiravir, ritonavir, lopinavir, remdesivir,\\
 & dexamethasone, ciprofloxacin, doxyciclin, oseltamivir, metronidazol, ivermectin  \\
PPE                  & mask, glove, gown, surgical, sanitiser, sanitizer, ppe  \\
Test                 &  test kit, covid test, pcr test, antigen test, corona test, diagnostic, diagnosis \\
Web domain           & https, www., http://, .com, .co.uk, .dk, .org, .info, .in, .net \\
\hline
\end{tabular}
    \caption{\textbf{COVID-19 products related keywords.} Keywords used to pre-select listings selling COVID-19 related products before their manual annotation, organised by category.}
    \label{tab:product_keywords}
\end{table}

Such method is not feasible as more products are searched, because the number of listings to be manually annotated is too large. As already done in our previous work~\cite{bracci2020covid}, we then limit our analysis to all listings mentioning COVID-19, using one the following keywords: ``corona virus'', ``covid'', ``coronavirus'' either in the title or description. To analyse COVID-19 related products, we first pre-selected a subset of these listings mentioning keywords in specific categories, see Table~\ref{tab:product_keywords}, and then manually annotated these listings. With respect to our previous effort, we find a new product category, which we call \emph{malware}, while no listing in the \emph{ventilator} category was found. Then, we characterize all listings mentioning COVID-19 by means of Natural Language Processing techniques. First, we apply BERT to build embeddings for the listings titles~\cite{devlin2018bert}, using the \emph{``paraphrase-mpnet-base-v2''}  implementation from the python package \emph{sentence-transformers}~\cite{reimers-2019-sentence-bert}   to map each title into 768-dimensional vectors.
We then use the UMAP algorithm~\cite{mcinnes2018umap} to map these  vectors in 2D, and cluster them using the \emph{hDBSCAN} algorithm~\cite{mcinnes2017hdbscan}. We then inspect each cluster by creating a ranking of words with tf-idf, and thus manually identify the product category each cluster refers to.

While each listing had an associated url to determine its uniqueness, which allowed us to track listing over time, vendors receiving bad reviews sometimes put identical copies of the same listing online. To overcome this issue and correctly count the number of listings, we created a new identifier of unique listings. We considered two listings as unique if the same vendor was posting listings in the same market, having only small variations in the title. We also excluded listings with prices larger than $40,000$ USD. Vendors post listings at high price to hold sales of these relative items, with the expectation of offering it again in the future.~\cite{soska2015measuring}.

\section*{Results}

\subsection*{Vaccine listings}

Our previous analysis~\cite{bracci2020covid} found 34 fake cures, including  antidotes, vaccines, and allegedly curative recreational drug mixes. These listings were scam, since no official vaccine was approved in the considered time period. Here, we  analyse COVID-19 listings since November 2020. We found 250 unique listings offering vaccines and manually categorised them in three categories: \emph{approved vaccines}, \emph{unspecified vaccines} and \emph{proofs of vaccination}. 
Listings in the \emph{approved vaccines} category explicitly mentioned official vaccines, an example being the Pfizer/BioNTech vaccine that was offered at 500 USD on the Invictus marketplace, see Figure~\ref{Approved_vaccine_Pfizer}. 
Listings in the \emph{unspecified vaccines} category instead referred to unbranded vaccines, for example by offering alleged unapproved vaccines well before official clinical trials were completed, as shown in Table~\ref{Other_vaccine_1}.
Listings in the \emph{proofs of vaccination} category offered a fabricated certificate of COVID-19 vaccination, as the fake COVID-19 passport offered at 55 USD on the Hydra marketplace, see Figure~\ref{Screenshots_vaccines} with its English translation in Table~\ref{Proof_of_vaccination_translation}.
The \emph{unspecified vaccines} category contained 94 listings, followed by the \emph{proofs of vaccination} category with 81 and then the \emph{approved vaccines} category with 75 listings. The \emph{unspecified vaccines} category also has the highest number of vendors, with 62 offering these products across 13 different DWMs. Similar statistics for the other categories can be found in Table~\ref{Category_Vaccine_listings}. 

\begin{figure}[h]
  \centering
  \includegraphics[width=15cm]{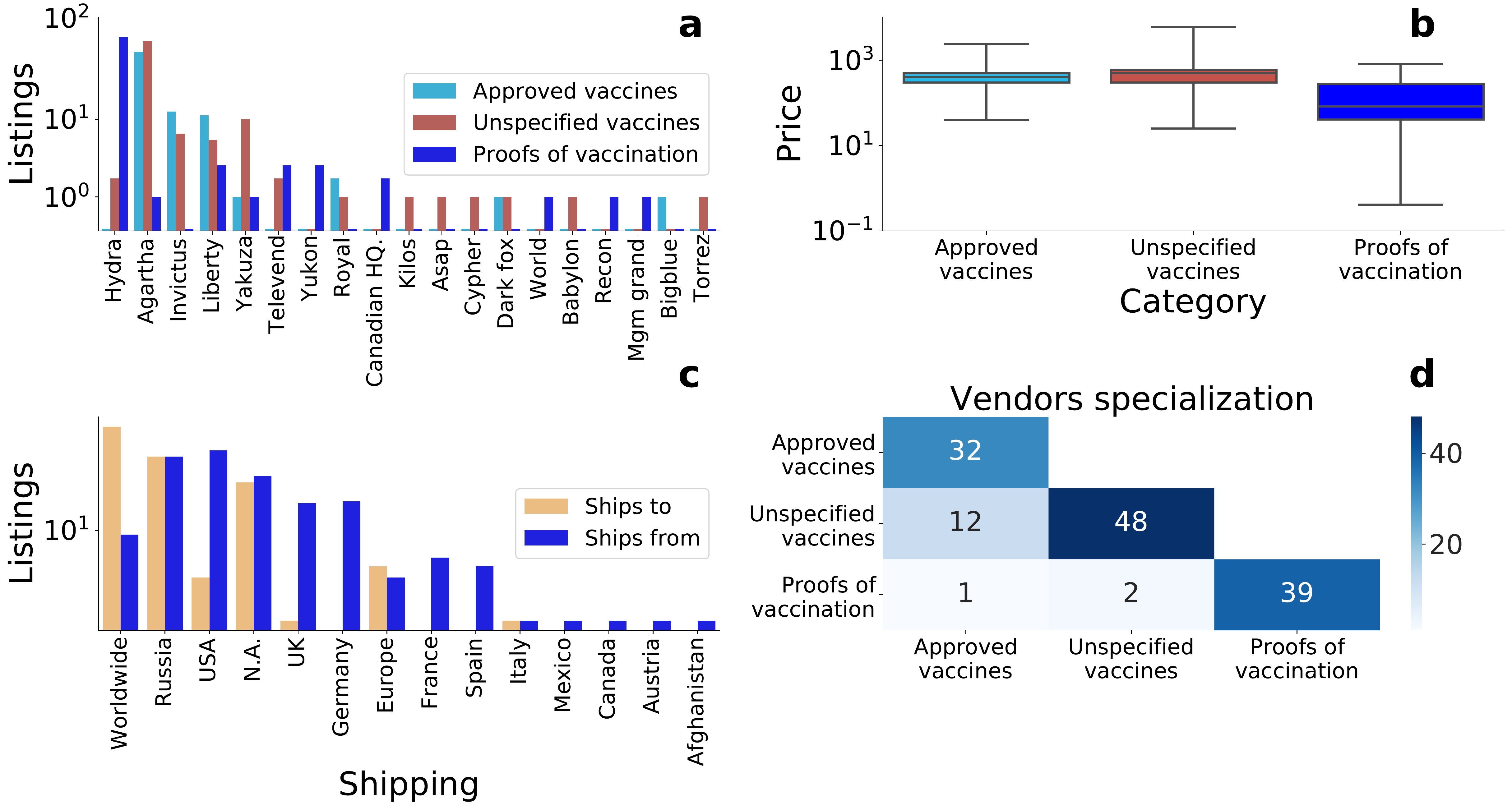}
  \caption{\textbf{DWMs and COVID-19 vaccines.} (a) Number of unique listings offered in each DWM. ``BB. house'' stands for Big brother house, while ``Canadian HQ.'' to The Canadian Headquarters. (b) Boxplots of the prices in USD at which vaccines were offered. Horizontal lines represent the median value, box ends the first and third quartiles, and whiskers minimum and maximum values, respectively. (c) Number of listings indicating where vaccines are declared to be shipped from and to. ``N.A.'' stands for not applicable and ``Russia'' for Russia and Eastern neighbouring countries. (d) Number of vendors offering a vaccine in a given category. Only the lower triangle of the matrix is shown because it is symmetric, where its diagonal represents vendors offering only listings in that category. }
  \label{Unique_listings}
\end{figure}

We now consider how the offer of vaccines was distributed across markets. The majority of vaccines were offered in the Agartha marketplace, with 108 listings, followed by Hydra with 67, which offered 65 out of the 81 fabricated COVID-19 vaccination certificates in our dataset. Figure~\ref{Unique_listings}(a) shows the category of listings offered by each DWM with at least one vaccine. 11 of these DWMs are specialized in offering only one category of listings, with one DWM only offering \emph{approved vaccines}, 5 DWMs only offering \emph{unspecified vaccines}, and 5 DWMs \emph{proofs of vaccination}. Three DWMs, Agartha, Liberty, and Yakuza, offer at least one listing in each of the three categories considered. The DWMs specialization can be seen in Figure~\ref{DWM_listings}. Vaccine listings have a short lifetime on a DWM, with most listings that are offered for less than 25 consecutive days, see Figure~\ref{Lifetime_vaccine_listings}. Such short lifetimes may be due to platform moderation, which in some cases explicitly prohibit such listings. However, such claims are not verifiable with our current data set.

Regarding the price of vaccine listings, Figure~\ref{Unique_listings}(b) shows its distribution in the three categories under consideration. Listings in the \emph{approved vaccines} category have prices ranging from 40 to 2,400 USD; listings in the \emph{unspecified vaccines} category between 25 USD to 6,060 USD; and listings in \emph{proofs of vaccination} category from less than 1 USD up to 814 USD. Proofs of vaccination were the cheapest products, probably because they consist of fake documentation (e.g., falsified COVID-19 passport). Price of \emph{approved vaccines} listings varied depending on the vaccine brand offered, see Figure~\ref{Unique_listings_price}. The first listing in this category to be offered was the Pfizer/BioNTech vaccine at 1,000 USD. The other 44 listings offering the Pfizer/BioNTech vaccine proposed prices ranging from 200 to 2,400 USD. The Astrazeneca/Oxford vaccine, the second to be officially approved, was offered on DWMs since December 27, 2020. Only four listings offered this vaccine, ranging from 300 to 900 USD. The other approved vaccines offered on DWM were Moderna with 21 listings, Johnson\&Johnson with four, Sputnik V with four, and Sinopharm with two. Their prices ranged from 40 to 2,000 USD. 

A natural next step is to analyse the geography of this trade, which we can do by looking at the shipping origin/destination information advertised in the listings. Most vendors declared that they would ship anywhere in the world, a behaviour that facilitates illicit trade. Vaccine warehouses were mostly in USA, followed by Germany and UK. Also, many listings do not declare any shipping information and all general shipping statistics are visible in Figure~\ref{Unique_listings}(c). In the 58\% of cases where no shipping information is declared, vendors invite potential customers to a direct interaction through Whatsapp, email, or phone. Percentage of listings where vendors suggest to initiate a direct interaction varies depending on the category considered. It happens for 78 (or the 84\%) of listings in the \emph{unspecified vaccine} category, for 64 (or the 85\%) of listings in the \emph{approved vaccine} category, and for only three listing in the \emph{proofs of vaccination} category. This last low number is due to Hydra marketplace, which sells 64 proofs of vaccinations but their vendors never shared their contact information. 

Do these vendors sell multiple kind of products related to vaccines? Or do they focus on a single category? Figure~\ref{Unique_listings}(d) shows that vendors offering \emph{proofs of vaccination} were specialised, with only one vendor also offering  \emph{approved vaccines} and two unspecified vaccines. On the contrary, 12 vendors were offering both \emph{vaccines} and \emph{unspecified vaccines}. We did not observe any vendor offering listings in all three categories. Moreover, most vendors offer only one COVID-19 listing and trade in only one DWM, with the notable exception of a vendor, who had twelve listings in eleven different DWMs, as detailed in Figure~\ref{histograms_vendors}. 

\begin{figure}[h]
  \centering
  \includegraphics[width=15cm]{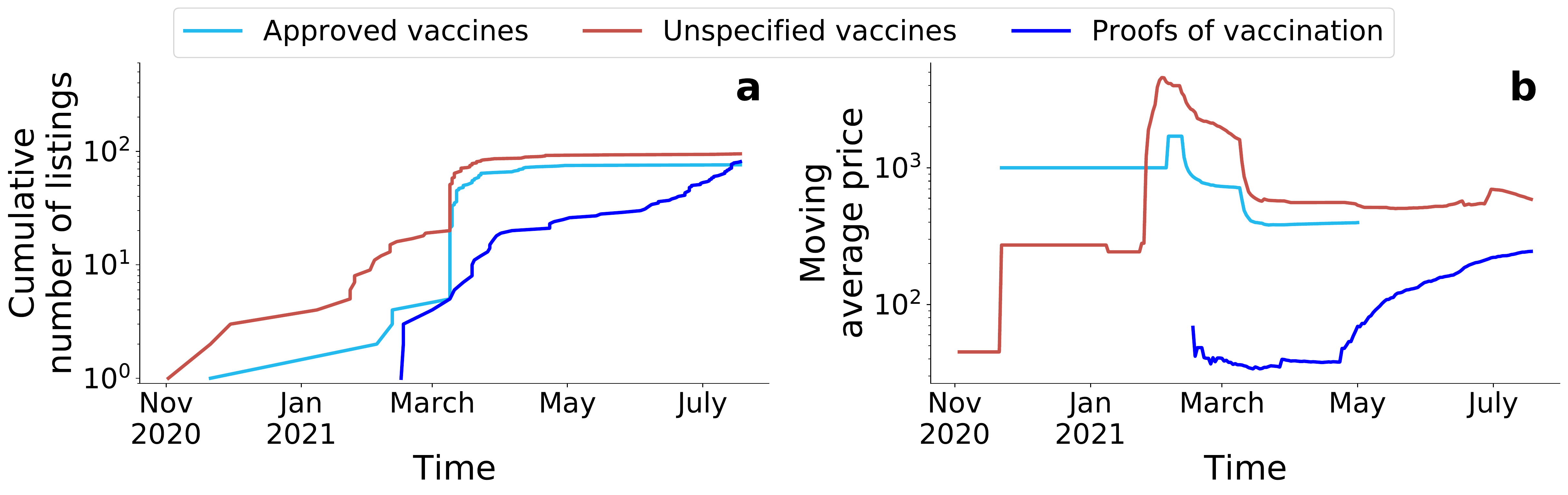}
  \caption{\textbf{Temporal evolution of COVID-19 vaccine listings.} (a) Cumulative number of listings over time in the three categories considered. (b) Average price over time in the same three categories, computed with a 90-days moving window.}
  \label{temporal_evolution_time}
\end{figure}

The evolution of vaccines on DWMs closely followed major COVID-19 related events, as shown in Figure~\ref{timeline_vaccines}. Figure~\ref{temporal_evolution_time}(a) shows that multiple vaccine listings were simultaneously present on DWMs when the first vaccination trials were undergoing, between March 16, 2020 and April 14, 2020~\cite{jackson2020mrna}. Toward the end of the first wave of contagions in Europe on June 9, 2020~\cite{lockdown_eased}, no more vaccine listings were present on DWMs from July 1, 2020. These listings reappeared on September 16, 2020, at the beginning of the second wave of infections that started in September 2020~\cite{second_wave_europe}. Up to this moment, we detected only COVID-19 listings in the \emph{unspecified vaccines} category. The first listing in the \emph{approved vaccines} category was a Pfizer/BioNTech vaccine and was offered since November 17, 2020, two weeks before its first official approval on December 2, 2020 by the UK~\cite{ledford2020uk}. A similar pattern was registered for the first AstraZeneca/Oxford vaccine listing on DWMs. It was offered on December 27, 2020, three days before the first official approval of this vaccine (by the UK) on December 30, 2020~\cite{UK_oxford_vaccine}. The remaining approved vaccines, Johnson\&Johnson, Moderna, Sputinik V, and Sinopharm, all appeared in the first half of March, when we started to monitor the Agartha marketplace. All \emph{approved vaccines} listings disappeared on DWMs since May 1, 2021, albeit there may be other DWMs offering these products that are not part of our analysis. Since listings in the \emph{unspecified vaccines} category continued to be observed until July 2021, we speculate that vendors were starting to have multiple vaccine brands, and they did not specify anymore which one are selling.  For more details, see Figure~\ref{temporal_evolution_approved_vaccines}(a). Listings in the \emph{proofs of vaccination} category emerged on February 15, 2021, when airlines were encouraging governments to allow certificates of vaccinations to become a way to safely travel~\cite{international_travel_1}.

Finally, we looked at the temporal evolution of the average price of these listings. The three categories followed different trends, as visible in Figure~\ref{temporal_evolution_time}(b). The price of \emph{unspecified vaccines} was high between March and May 2020, when DWMs vendors likely tried to profit from the initial lack of COVID-19 medications~\cite{bracci2020covid}. Afterwards, their mean price has gradually decreased, meaning that the new listings appearing on DWMs were offered at progressively lower prices. However, the average price rose back to March levels in January, when vaccinations campaigns around the world were starting. The availability of officially tested vaccines led to the emergence of relative listings on DWMs since November 2020. The average price of these listings have floated over time between a few hundreds USD to more than a thousand. For more details, see Figure~\ref{temporal_evolution_approved_vaccines}(a). Finally, the needs for a certificate of COVID-19 vaccination had meanwhile increased, and so had the price of listings in the \emph{proofs of vaccination} category. 
 Vaccines certificates have gradually become mandatory in many countries, and especially for international travel, and their sale on DWMs confirms what researchers had hypothesised~\cite{brown2020passport}, warning against similar situations happening in the future.   

\subsection*{Other COVID-19 related products}

\begin{figure}[h]
  \centering
  \includegraphics[width=15cm]{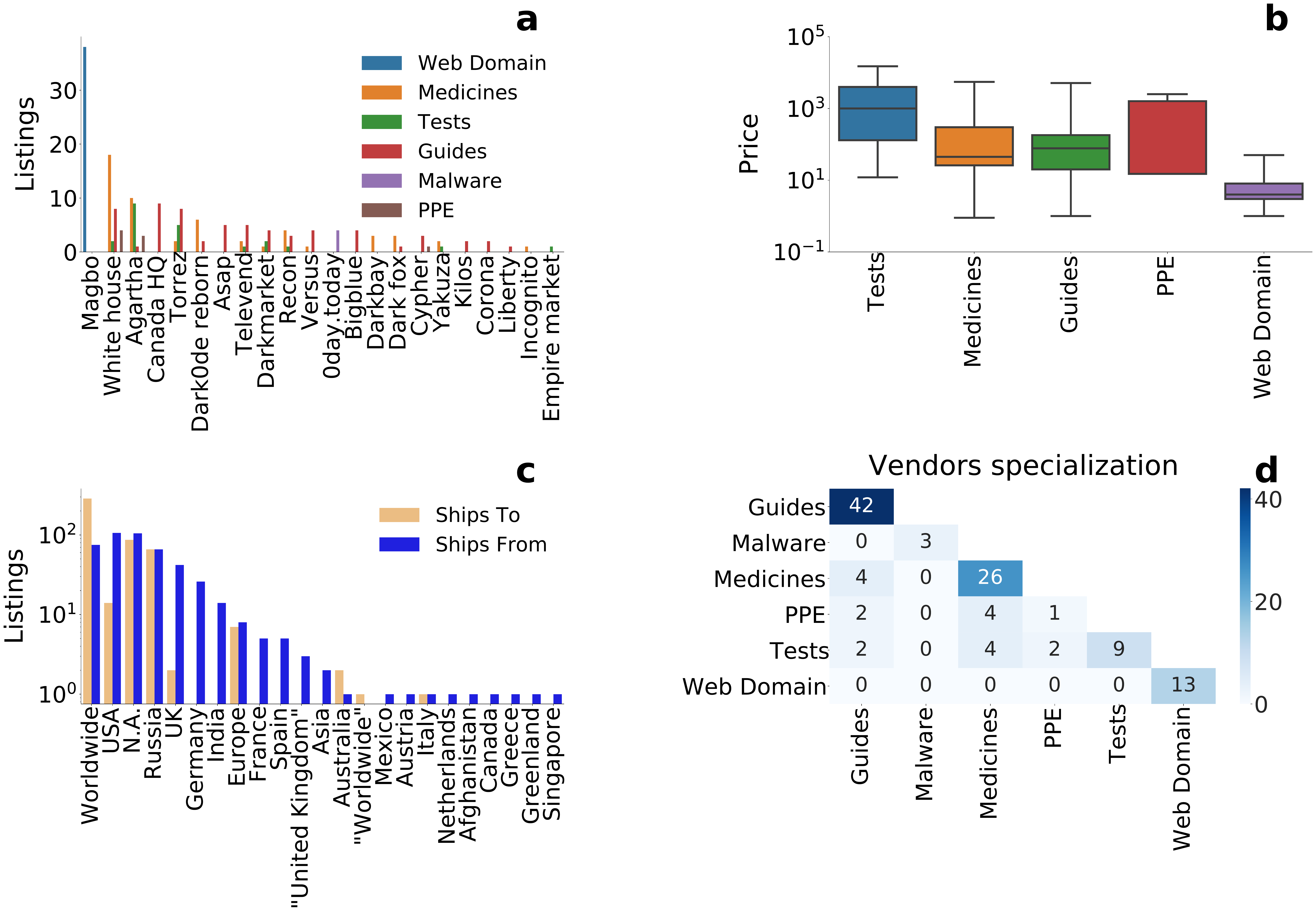}
  \caption{\label{fig:covid_products} \textbf{COVID-19 related products.} (a) Break-down of COVID-19 related products by category and market. (b) Boxplot representing the price distribution of listings in each category. Horizontal lines represent the median value, box ends the first and third quartiles, and whiskers minimum and maximum values, respectively. (c) Number of listings indicating where COVID-19 related products are declared to be shipped from and to. ``N.A.'' stands for not applicable and ``Russia'' for Russia and Eastern neighbouring countries. (d) Number of vendors offering a COVID-19 related product in a given category. Only the lower triangle of the matrix is shown because it is symmetric, where its diagonal represents vendors offering only listings in that category.}
\end{figure}

DWMs have been a venue for the sale of other licit and illicit COVID-19 related products, like PPEs, tests, or medicines, as reported for the period from January to November 2020~\cite{bracci2020covid}. Here, we monitor COVID-19 related products in the second part of the pandemic, between November 2020 and July 2021, see Table~\ref{tab:covid_19_products} and Figure~\ref{fig:covid_products}. Listings are divided in six different categories: \emph{PPEs} represent healthcare objects like masks; \emph{medicines} COVID-19 related medicines like hydroxychloroquine; \emph{guides on scamming} are instructions on how to get relief funds; \emph{tests} represent COVID-19 tests; \emph{web domains} that are related to COVID-19 like "covidtest4you.com"; and \emph{malware} represents malicious software to hack COVID-19 test or vaccination records software. Listings from these categories are offered in 21 DWMs, and are available in multiple markets. \emph{Malware} and \emph{web domains} are an exception because sold in two specific markets only.
We find that \emph{PPEs} and \emph{medicines} have almost disappeared from DWMs  w.r.t. previous observations  ~\cite{bracci2020covid,broadhurstavailability}. \emph{PPEs} listings are mostly advertising bulk sales at high prices, coherently with the end of shortages of these products, while \emph{medicines} listings, like hydroxychloroquine, are substituted by vaccines and present on DWMs in a lower number, with only 3 listings advertising Ivermectin~\cite{ivermectin}. On the contrary, \emph{guides on scamming} were still present with comparable numbers, claiming to teach ways to access COVID-19 relief funds in different countries. Notably, the number of listings offering COVID-19 tests had also increased, with tests increasingly being required for travel or work. We also found 4 listings advertising malware to illicitly access official systems to record test results or even vaccinations.

Figure~\ref{fig:covid_products}(a) shows the distribution of unique listings on each DWM offering them. These listings are concentrated in 4 DWMs, with the majority of them offering less than 5 listings. However, there is less category specialization w.r.t. what observed for the vaccines, with multiple markets offering listings in different categories. Prices are also very heterogeneous, with \emph{test}'s median price highest at 1000 USD, and \emph{web domain}'s lowest at just 4 USD, but also listings inside the same category ranging from 10 USD to 1000 USD in all categories but \emph{web domain}, see Figure~\ref{fig:covid_products}(b).
In Figure~\ref{fig:covid_products}(c), we show the origin and destination of the considered listings, as declared by vendors. The majority of listings declare to be shipping worldwide, while the United States is the country appearing the most as declared origin of the listings. Russia and Eastern neighbouring countries are both origin and destination, mainly because of \emph{proof of vaccination} listings offered on Hydra, whereas UK, Germany and India appear almost only as countries of origin. Other countries/regions are declared, but less frequently.
Figure~\ref{fig:covid_products}(d) shows vendors specialization regarding COVID-19 related products. All categories show highly specialised vendors, except \emph{PPE}, where only one vendor out of seven sells only in that category, and \emph{tests}, where less than 55\% of vendors sell only such products.

\subsection*{Listings with COVID-19 mentions}

As previously done in our first analysis~\cite{bracci2020covid}, we assess the overall impact of COVID-19 on DWMs, through the analysis of all listings mentioning COVID-19 either in the title or body. We extend our previous analysis by considering listings appearing until July 2021 and by categorising the selected listings, providing a richer and deeper picture of how DWMs were indirectly affected by COVID-19. 

\begin{figure}[h]
  \centering
  \includegraphics[width=15cm]{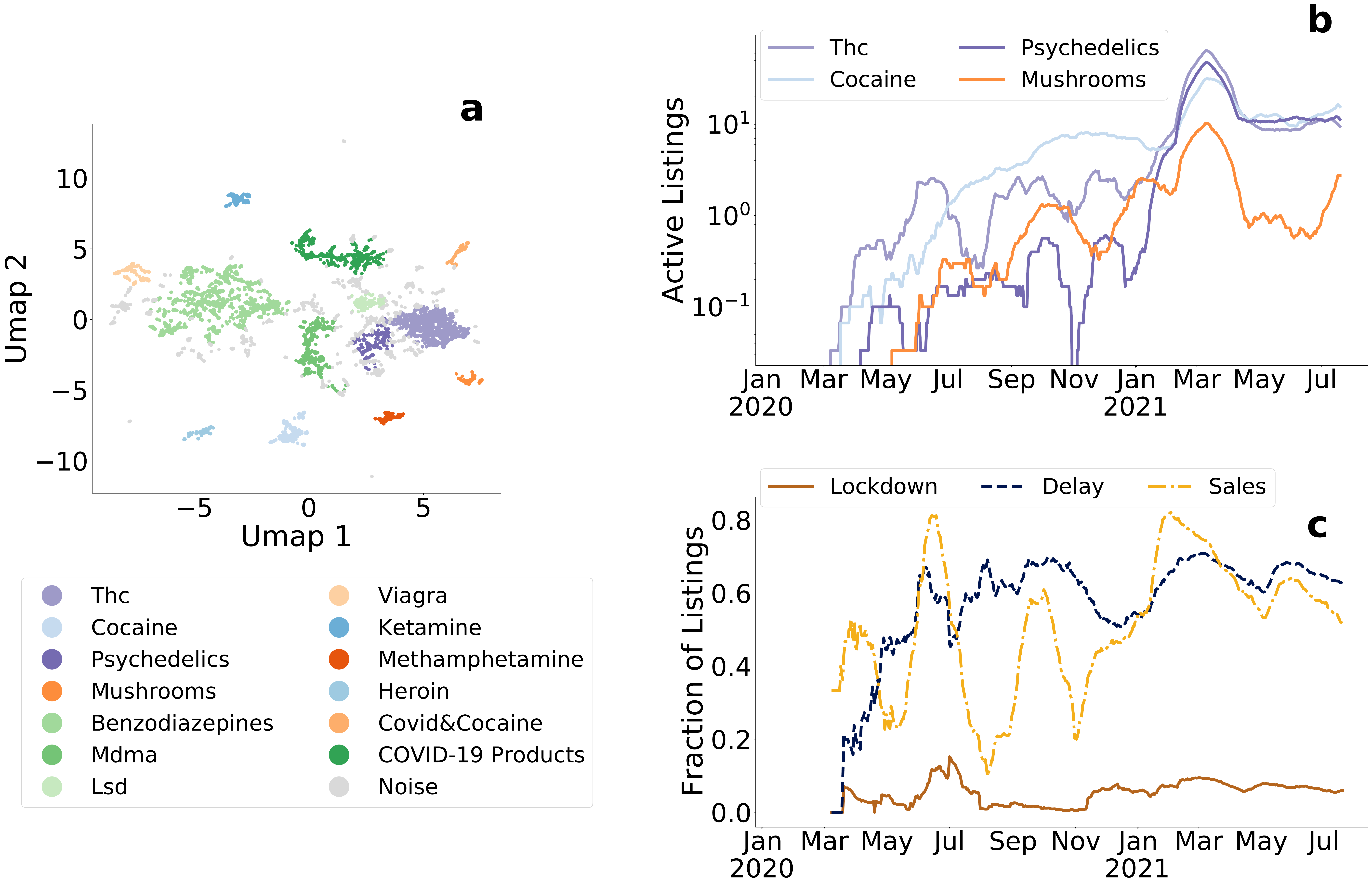}
  \caption{\label{fig:topic_modelling} \textbf{Characterization of COVID-19 mentions.} (a) UMAP representation of doc2vec embeddings. hDBSCAN clustering finds 13 meaningful categories covering COVID-19 related products and all major drugs sold on DWMs. (b) 30-day rolling average of active listings in 4 categories of listings mentioning COVID-19: \emph{thc}, \emph{psychedelics}, \emph{cocaine}, and \emph{mushrooms}. (c) 30-day rolling average of fraction of previously identified drugs listings mentioning 3 different COVID-19 related themes: lockdowns, shipping delays and sales.}
\end{figure}

We characterise products mentioning COVID-19 with state-of-the-art Deep Learning based Natural Language Processing techniques~\cite{devlin2018bert, mcinnes2018umap, mcinnes2017hdbscan}, see Methods for more details.  
As shown in Figure~\ref{fig:topic_modelling}(a), we find 13 different categories of listings corresponding to different kind of products. In addition to the already discussed COVID-19 related products, only drugs appear to be mentioning COVID-19, while other traditional DWMs' products like stolen IDs or credit card dumps don't, showing which kind of goods reacted to, or where affected by, the pandemic. We then analyse the temporal evolution of these categories. We show the number of active listings for 4 large categories in Figure~\ref{fig:topic_modelling}(b), while all other categories are shown in Figure~\ref{fig:appendix_topic_modelling}. Drugs show an overall increasing trend throughout the whole period. Different categories, however, display different fluctuations in time, showing how different goods behave in an heterogeneous way with respect to COVID-19. For example, at the end of our covered period we can see \emph{thc} and \emph{psychedelichs} showing a flat trend, while \emph{cocaine} and \emph{mushrooms} are increasing. 

While it is not possible to understand the reasons behind each single temporal trend, we can gain more insights on why drugs are increasingly mentioning COVID-19 by investigating which themes are recurrent in these listings. In Figure~\ref{fig:topic_modelling}(c), we analyse mentions of three different themes which have been previously observed in DWMs~\cite{bracci2020covid}: \emph{lockdown}, \emph{sales}, and \emph{delay}, and which can be used as general proxies of the indirect impact of the pandemic. \emph{Lockdown} mentions are always lower than the other two themes, peaking in summer 2020 but staying always lower than $20\%$. \emph{Delay} mentions instead rapidly increase during the first months of the pandemic, and have been oscillating around $60\%$ of the listings since then, showing how drug vendors have been warranting possible delays throughout the whole observed period, confirming what's already been independently shown for the first phase of the pandemic~\cite{shipping_problems_dwm}. Finally, \emph{sale} mentions show larger fluctuations between as low as $15\%$ to even $80\%$. In particular, we can observe peaks related to the pandemic at the beginning of key COVID-19 related events: lockdowns in March/April 2020, in Summer 2020 coincidentally with openings in the western world, in October 2020 when the second wave started hitting Europe, and in February 2021 when the Delta variant started spreading in the world. By looking at mentions of these themes across all listings in our dataset, we find that overall mentions of \emph{lockdown}, \emph{sales}, and \emph{delay}, have decreased since the beginning of the pandemic, validating our finding that drugs-related listings are the product most impacted by COVID-19, see Figure~\ref{fig:overall_mentions}.

Automatic keyword search has allowed us to uncover macroscopic trends, but it fails to capture finer details which can only be uncovered by in-depth looks at the texts of the listings. We therefore resort to a qualitative analysis of their descriptions. First, we already noticed that mention of delays in drug listings are still frequent, amounting to 56 \% of the listings. While vendors generally preemptively mention possible delays due to COVID-19, we find numerous mentions of USA based vendors blaming UPS for these, as shown in one example reported in Table~\ref{tab:example_usps}: "THE USPS IS UNDERFUNDED AND MAY BECOME UNRELIABLE COMPARED TO THE PAST! (ESPECIALLY DURING COVID-19 AND HOLIDAYS!)". These claims reflect widely reported issues with the United States Postal Services since June 2020~\cite{usps_crisis}.
Moving away from delays, we find that ~10\% of vendors mention COVID-19 by ensuring potential clients that they are taking all necessary safety measures when preparing the deliveries. An example of this is reported in Table~\ref{tab:drug_listing_staysafe}. Finally, we find listings mentioning limited stocks due to the pandemic, as shown in Table~\ref{tab:example_drug_listing_stocks}, where the vendor claims that``stocks are almost exhausted by Corona Covid 19''.

\section*{Discussions and conclusion}

In this paper, we have extended previous analyses~\cite{bracci2020covid,broadhurstavailability} on the impact of COVID-19 on DWMs both in terms of duration of the monitored period and of breadth of the analysed products. The covered period, ending at the end of July 2021, included the second phase of the pandemic, i.e., when vaccines become available. We have identified a sharp increase on the number of listings selling vaccines and proofs of vaccinations, from 34 between January and November 2020  (when no vaccine had been released yet)   to 250 after, including officially approved vaccines like Pfizer or Moderna. Vaccine related listings have replaced other previously observed COVID-19 related products (e.g. PPE and hydroxychloroquine), whose presence has been steadily decreasing since November 2020. While assessing the overall COVID-19 impact through the analysis of listings explicitly mentioning COVID-19, we have found that drugs were the most affected traditional DWMs product.

A key contribution of the present work is the study of the interplay between DWMs and the COVID-19 pandemic, after the official approval of vaccines.
It was previously shown that, when a product is in shortage in the regular economy, or public attention is focused on it, listings advertising its sale appear on DWMs. For instance, this is what happened for PPE and hydroxychloroquine during the first phase of the pandemic. Since  in the observed period   these products were easily available on regulated markets, we coherently detect that these products disappeared in the second phase of the pandemic. In late 2020, we have seen the same pattern with vaccines, which appeared around the time of their official approval, reflecting the claims of other mass media news~\cite{vaccines_covid_DWM1, vaccines_covid_DWM2, vaccines_covid_DWM3}. They then spiked at the beginning of 2021, to be later replaced by fabricated proofs of vaccinations with the increasing need of vaccine passports and green passes~\cite{Green_pass,Green_pass_france}. Mentions of lockdowns, delays, and sales followed a similar dynamics, with spikes observed in the first phase of the pandemic in 2020 and their mentions steadily decreasing during the second phase. However, we found that drugs listings mentioning COVID-19 increased in time, with numerous mentions of delays and sales, some of which are related to stock shortages and increase in health security measures, as unveiled by our qualitative analysis.

A limitation of the present work is that, while the number of DWMs simultaneously monitored over time is greater than most previous studies, we cannot ensure that all DWMs were surveyed. In fact, the number of active DWMs is constantly changing due to closures or new openings~\cite{elbahrawy2019collective} and obtaining full coverage is challenging due to the active efforts of DMWs to obstruct research studies and law enforcement investigations. Moreover, our study is limited to what vendors advertise on these platforms, as we have no data on actual purchases to quantify how many people have been endangered by this phenomenon. Future work, relying on backend servers seized during police takedowns of DWMs, could improve on our study by overcoming these limitations.

 The diffusion of illicit vaccines on DWMs, the simultaneous decrease of PPEs and medicines, confirms the link between product shortages, public attention and listings on DWMs. 
This phenomenon has the potential to pose a serious public health threat, as DWMs have become increasingly easier to access, resilient to police closures~\cite{elbahrawy2019collective} and shown to be a catalyst for decentralized peer to peer trading between buyers and sellers of illicit items~\cite{nadini2022emergence}.
The purchase of unregulated, and possibly fraudulent, health related items on DWMs poses a concrete health risk for the buyers. 
Moreover, the availability of fake vaccination or testing record risks to undermine public health measures implemented by numerous countries worldwide, and calls for more investigation of this phenomenon for the current and future pandemics.
By highlighting how vaccines appear on DWMs, our analysis may help raise awareness of the phenomenon and support the effort of law enforcement agencies to contain it by repeating past successful approaches~\cite{Operation_Onymous, FBIAlphabay, WallStreetMarket, SizedBerlusconiMarket, BillionFedsSilkRoad, JokerStashDisruption2020, SILKROADSEALED}.
Furthermore, our results call for more investigation of DWMs to anticipate such dangers in future public health crisis.  


\section*{Competing interests}
  The authors declare that they have no competing interests.

\section*{Author's contributions}

All authors designed the research. M.A. and I.G. provided the data. A.Br., M.N., M.A. and I.G. pre-processed and analysed the data. All authors analysed the results and wrote the manuscript. All authors read and commented on the manuscript. A.Ba. coordinated the project. 

\section*{Acknowledgements}
A.Br., M.N., A.T., A.G and A.Ba. were supported by ESRC - Economic and Social Research Council as part of UK Research and Innovation’s rapid response to COVID-19, through grant ES/V00400X/1. M.A and D.M., acknowledge support from the U.S. National Science Foundation grants 1717062 and 2039693.

\section*{Correspondence}
Correspondence should be addressed to \\Andrea Baronchelli: abaronchelli@turing.ac.uk.


\newpage
\appendix

\section*{Examples of listings detected}
\label{Examples_of_listings}
  
\begin{figure}[h]
  \centering
  \includegraphics[width=11cm]{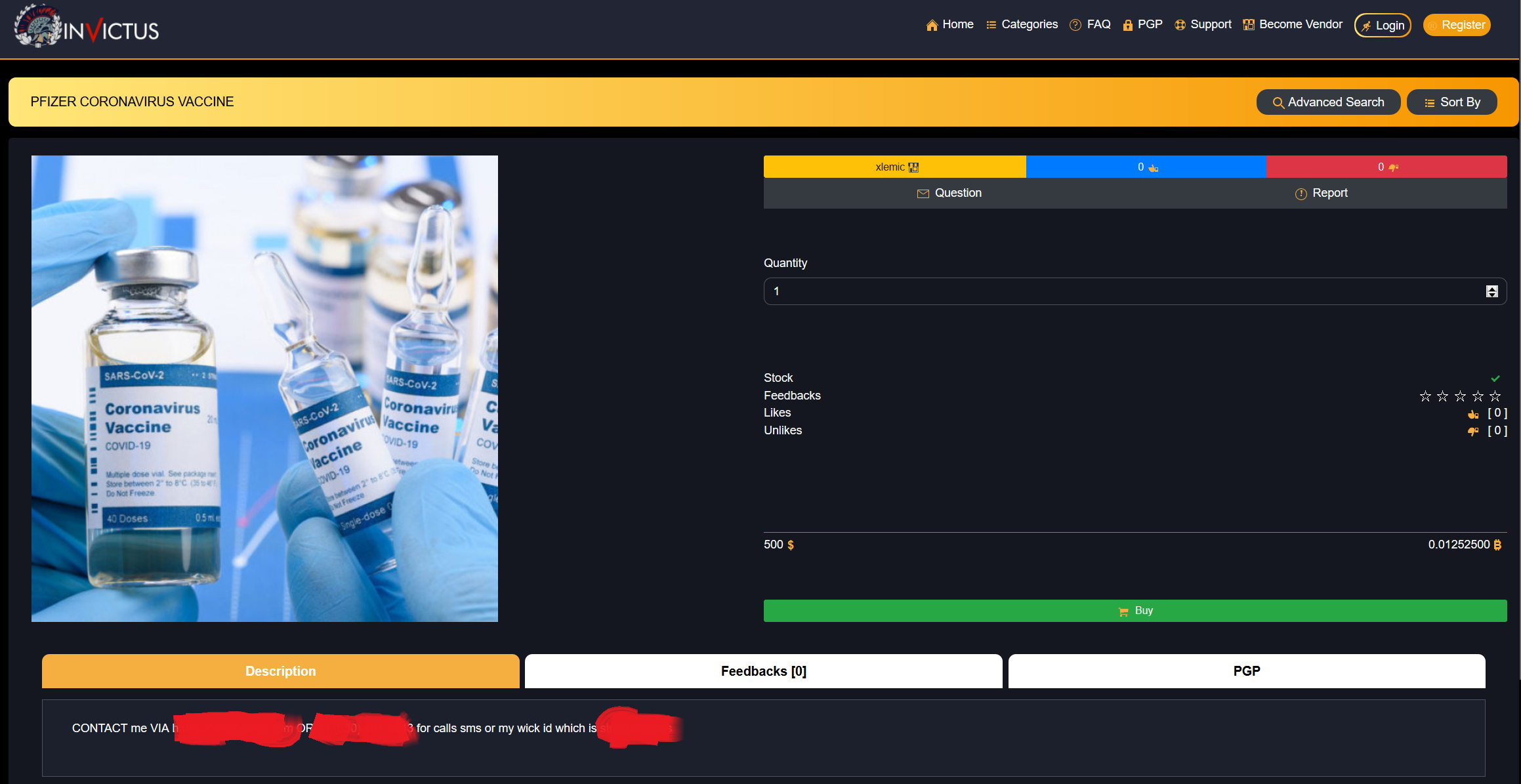}
   \caption{\textbf{Pfizer/BioNTech vaccine offered on Invictus.} Screenshots of a listing in the \emph{approved vaccines} category offering the Pfizer/BioNTech vaccine at \$500 on the Invictus marketplace. We removed the contact information of the vendor, who invites the potential customer to have a direct contact. The screenshot was taken on February 6, 2021.}
  \label{Approved_vaccine_Pfizer}
\end{figure}

\begin{table}[h]
\caption{\textbf{Generic vaccine offered on a DWM.} Example of a listing in the \emph{unspecified vaccines} category offering a generic vaccine, which does not specify the producer. Personal information of the vendor are hidden with \# symbols.}
      \begin{tabularx}{\textwidth}{l|X}
        \hline
        Title & \thead{COVID-19 antidote from china. offering at 15k USD} \\ 
        \hline
        Body & {\footnotesize the covid-19 current massacre is supposed to have ended by now. while the who is trying to be selfish with human life, we are trying to save the lives. the real virus is the leaders. this vaccine should be used just once on one person and basically the giveaway price i put here is nothing compare your life. get your vaccine now in time. you can buy from me and resell at your price. contact me for more details. email: \#\#\#\#\# wickr: \#\#\#\#\# telegram: \#\#\#\#\# kik: \#\#\#\#\# } \\ 
        \hline
        Price & \thead{15,000 USD}\\
        \hline
        Shipping from/to & \thead{N.A.} \\
        \hline
        Vendor & \thead{\#\#\#\#\#} \\  
         \hline
        DWM & \thead{DarkBay} \\
      \end{tabularx}
\label{Other_vaccine_1}
\end{table}

\begin{figure}[h]
  \centering
  \includegraphics[width=11cm]{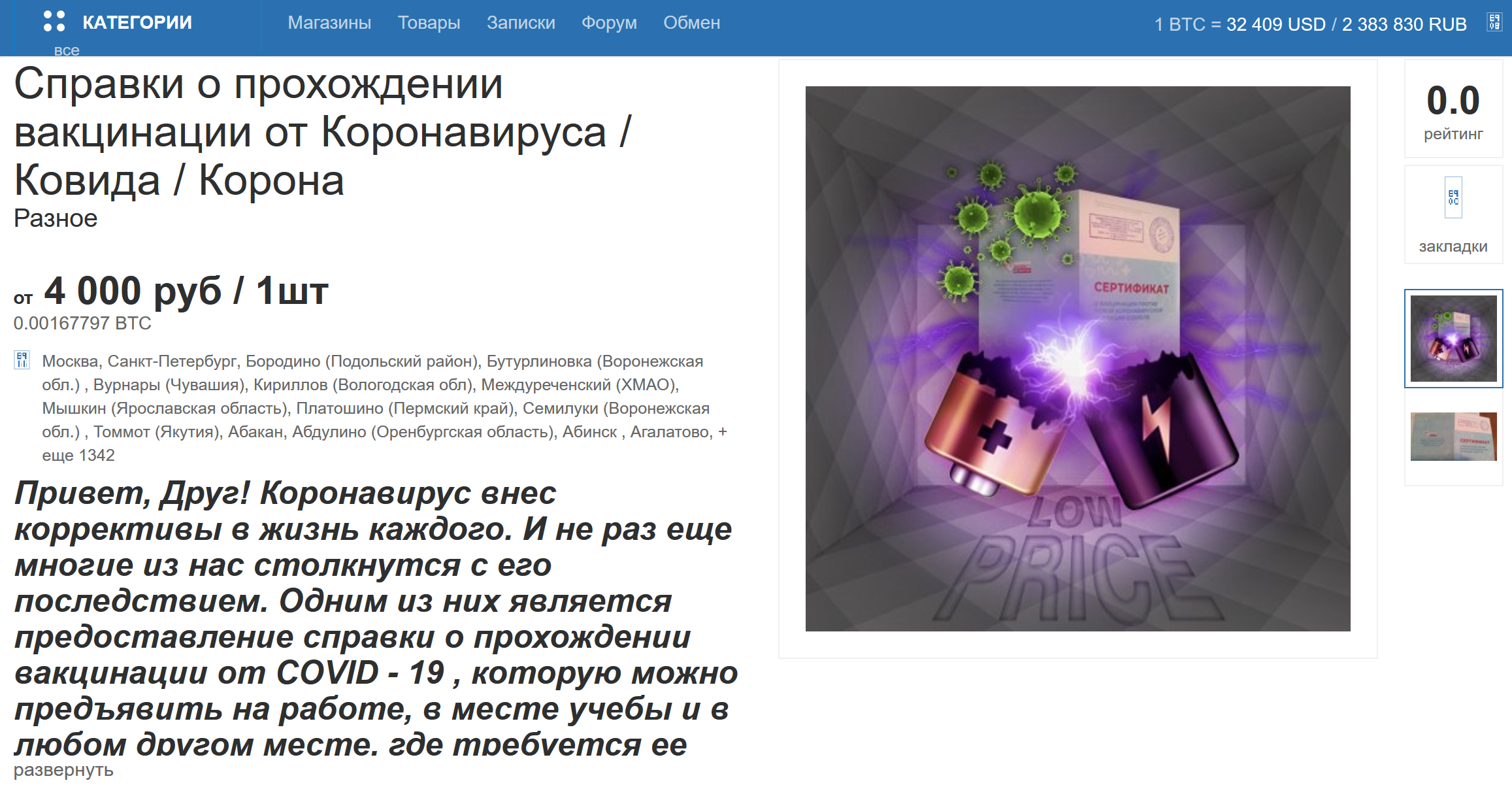}
   \caption{\textbf{Proof of vaccination offered on Hydra.} Proof of vaccination offered at 4,000 rubles (55 USD) detected on the Hydra marketplace. The original language of this listing is Russian and its translation in the English language is given in Table~\ref{Proof_of_vaccination_translation}. The screenshots were taken on February 6, 2021.}
  \label{Screenshots_vaccines}
\end{figure}

\begin{table}[h]
\caption{\textbf{Translation of the proof of vaccination offered on a Hydra.} English translation of the listing in Figure~\ref{Screenshots_vaccines}(b). We use Google Translator to translate the text from Russian to English.}
      \begin{tabularx}{\textwidth}{l|X}
        \hline
        Title & \thead{Coronavirus / Covid / Corona vaccination certificates} \\ 
        \hline
        Body & {\footnotesize Hello Friend! The coronavirus has made adjustments to everyone's life. And more than once many of us will face its consequences. One of them is the provision of a COVID-19 vaccination certificate, which can be presented at work, at the place of study and in any other place where it is required. To order, you need to indicate your full name, date of birth, date of the issued document. If you need a non-Moscow institution, you will need to pay extra for the production of the necessary seals. Production time is 2-7 days. Delivery when sent by courier will have to be paid for upon receipt. It is possible to send by registered or regular mail, then we will take on this heavy burden. An ordinary letter has no track, therefore, until we receive the letter, we will remain in the dark about its fate.} \\
        \hline
        Price & \thead{55 USD}\\
        \hline
        Shipping from/to & \thead{Russia and neighbouring Eastern countries/Russia and neighbouring Eastern countries} \\
        \hline
        Vendor & \thead{\#\#\#\#\#} \\
        \hline
        DWM & \thead{Hydra} \\
      \end{tabularx}
\label{Proof_of_vaccination_translation}
\end{table}

\begin{table}[h]
\caption{\textbf{Example of drug listing mentioning COVID-19, and problems with the United States Postal Services (USPS).}}
      \begin{tabularx}{\textwidth}{l|X}
        \hline
        Title & \thead{(10) 30mg adderall pressed pills: us - us} \\ 
        \hline
        Body & {\footnotesize
Adderall is used in the treatment of attention deficit hyperactivity disorder (ADHD) and narcolepsy. It is also used as an athletic performance enhancer, cognitive enhancer, appetite suppressant, and recreationally as an aphrodisiac and euphoriant. It is a central nervous system (CNS) stimulant of the Phenethylamine class.

You MUST Be A Minimum Of 18 Years of Age

For Research Purposes Only
Not For Human Consumption    Refund Policy  ALL SALES ARE FINAL! 

REFUNDS WILL NO LONGER BE ISSUED DUE TO SCAMMING! RESHIPS ARE ALWAYS AVAILABLE ON A CASE BY CASE BASIS, AND USUALLY ONLY WHEN A TRACKING NUMBER NEVER ORIGINALLY SCANS, OR GETS STUCK FOR 15+ DAYS. 

NO reships will be sent in the event of a tracking status of RETURN TO SENDER or UNDELIVERABLE AS ADDRESSED.

Reships DO NOT qualify if a package status is marked as Delivered or indicates the package is In Transit to its destination. If the package is in the system, please wait it out for the package to arrive. THE USPS IS UNDERFUNDED AND MAY BECOME UNRELIABLE COMPARED TO THE PAST! (ESPECIALLY DURING COVID-19 AND HOLIDAYS!)

Use a real name and address for your package. If a package is stuck IN TRANSIT for a few days, and a tracking number is given to you, please call USPS to locate it. THOUGH I MAY CARRY NON SCHEDULED RESEARCH CHEMICALS, DO NOT CLAIM TO KNOW THE CONTENTS. NO PACKAGES WILL EVER REQUIRE A SIGNATURE!

} \\
        \hline
        Price & \thead{59.13 USD}\\
        \hline
        Shipping from/to & \thead{USA/USA} \\
        \hline
        Vendor & \thead{\#\#\#\#\#} \\
        \hline
        DWM & \thead{Dark0de Reborn} \\
      \end{tabularx}
\label{tab:example_usps}
\end{table}

\begin{table}[h]
\caption{\textbf{Example of drug listing mentioning COVID-19, and ensuring safety measures are taken.}}
      \begin{tabularx}{\textwidth}{l|X}
        \hline
        Title & \thead{black diamond og sfv og shake popcorn        thc} \\ 
        \hline
        Body & {\footnotesize thank you so much for shopping with us  we are confident you'll love your order  while your here  take a moment to browse through our vendor page to see the many great strains  bulk orders  and emeraldgallipot promotional offers we have to offer 

   what we offer
   fast communication  all msg are answered within    hrs 
   fast delivery  product will be shipped on the next business day after order confirmation via usps priority mail 
   stealth packaging  vacuum sealed  odorless  sterile packaging 
   package tracking  available upon request three days after order confirmation 
   full refund replacement if tracking confirms package seized lost

   what we ask
   please provide your full address immediately in pgp format in buyer's note  use your full name and double check your address  deliveries that tracking confirms lost because of errors in provided information are not available for refund or replacement 

    all shipping addresses must be in the following format 
    name              john doe
    address              nameless ln
    city state zip  city xx      

   please finalize asap upon receiving package
   please leave a positive rating  if you are unhappy with your order  please tell us  we are happy to work with you to satisfy your needs

strain highlights
black diamond og
indica dominant hybrid
backberry kush diamond og
thc         

flavor   aroma
a cross between blackberry and diamond og  its flowers have a glittery trichome covering
and purple coloring that make it a beautiful gem to look at  the strains aroma is musky and earthy  almost like a deep red wine 

euphoric effects
black diamond is known to cause fits of giggles and is a great strain for hanging out with friends and creative expression 

medical benefits
ideal for patients who need strong medication but still want to be active and sociable  this strain tends to make consumers extremely hungry  making it a good choice for those looking to increase their appetite  just make sure you have some snacks on hand

san fernando valley og
sativa 
og kush direct
thc         

flavor   aroma
sfv og by cali connection is a sativa dominant hybrid that is as the name indicates  this og kush relative originates from californias san fernando valley  although their names are barely distinguishable  sfv og kush is actually the afghani crossed child to sfv og  leading with aromatic notes of earthy pine and lemon 

euphoric effects
creates a long lasting head haze and full body effect that leaves you feeling happy and relaxed  without damping your energy 

medical benefits
great for patients who need strong pain relief but dont want to be stuck on the couch 

note  we here at the emeraldgallipot take our customers safety as our highest priority  and to help protect you against the spread of the coronavirus  all packages we send are being thoroughly sterilized with a mild disinfectant and bleach solution prior to shipping for your protection  stay safe out there} \\
        \hline
        Price & \thead{50 USD}\\
        \hline
        Shipping from/to & \thead{USA/USA} \\
        \hline
        Vendor & \thead{\#\#\#\#\#} \\
        \hline
        DWM & \thead{Torrez} \\
      \end{tabularx}
\label{tab:drug_listing_staysafe}
\end{table}

\begin{table}[h]
\caption{\textbf{Example of drug listing mentioning COVID-19, and ``stocks are almost exhausted by Corona''.}}
      \begin{tabularx}{\textwidth}{l|X}
        \hline
        Title & \thead{grams speed paste normal quality} \\ 
        \hline
        Body & {\footnotesize When you place an order you agree with our conditions!

Offer: 5 Grams Speed Paste Normal Quality

This product is made from high grade washed A-Oil

Purity: 45\% up to 55\%

Approximately 20\% of the weight is lost during the drying process

For any questions feel free to contact us, we are happy to help you!

--------------------------------------------------------------------------------------------

Welcome to \#\#\#\#

The best speed (amphetamine) products on the market!
We sell from the normal quality till the highest quality you can get!
Our sending fits every mailbox!
We ship from Monday till Friday!

We ship from Germany and we know how to ship!
It is important for us that all orders arrive in all safety!

SHIPPING TIME
Europe:        	 2 to 7  Business days
Worldwide:       4 to 20 Business days 

REFUND and RESHIP
If orders not arrive please send us a message and we find a solution.
In case of non-arrival, we will reship 50\% or a 50\% refund.
Mistakes made in the address-format we will never reship or refund.
New buyers without any order history we never refund or reship.

Please give us some great feedback if you are happy with us!

AmphetamineCowboys

--------------------------------------------------------------------------------------------  UPDATE 13-02-2021

Dear customers,
From today 13-02-2021 we will go into vacation mode for 10 days until 23-02-2021. We do this because we have a lot of money in escrow and our stocks are almost exhausted by Corona Covid 19. New stocks are on the way but unfortunately it is slowing down due to Covid bullshit. We do not want to disappoint. We will receive new stocks next week so that we can continue on 23-02-2021. Of course all accepted orders have been shipped including today! We are online every day for all your questions about the shipped orders or for any other questions. Hoping for some understanding from you, we will be back soon on 23-02-2021. All be safe and hope to see you soon!

Sehr geehrte Kunden,
Ab heute 13.02.2021 werden wir fr 10 Tage bis zum 23.02.2021 in den Urlaubsmodus wechseln. Wir tun dies, weil wir viel Geld im Treuhandkonto haben und unsere Aktien von Corona Covid 19 fast erschpft sind. Neue Aktien sind auf dem Weg, aber leider verlangsamt sie sich aufgrund von Covid-Bullshit. Wir wollen nicht enttuschen. Wir werden nchste Woche neue Aktien erhalten, damit wir am 23.02.2021 weitermachen knnen. Natrlich wurden alle angenommenen Bestellungen auch heute noch versendet! Wir sind jeden Tag online fr alle Ihre Fragen zu den versendeten Bestellungen oder fr andere Fragen. In der Hoffnung auf ein Verstndnis von Ihnen werden wir bald am 23.02.2021 zurck sein. Alle sind in Sicherheit und hoffen, Sie bald zu sehen!

Chers clients,
 partir d'aujourd'hui 13/02/2021, nous passerons en mode vacances pendant 10 jours jusqu'au 23/02/2021. Nous faisons cela parce que nous avons beaucoup d'argent en squestre et que nos actions sont presque puises par Corona Covid 19. De nouvelles actions sont en route mais malheureusement, elles ralentissent  cause des conneries de Covid. Nous ne voulons pas dcevoir. Nous recevrons de nouveaux stocks la semaine prochaine afin de pouvoir continuer le 23/02/2021. Bien sr, toutes les commandes acceptes ont t expdies, y compris aujourd'hui! Nous sommes en ligne tous les jours pour toutes vos questions sur les commandes expdies ou pour toutes autres questions. En esprant une comprhension de votre part, nous serons de retour bientt le 23/02/2021. Soyez tous en scurit et esprons vous voir bientt!

Welcome to \#\#\#\#

The best speed (amphetamine) products on the market!
We sell from the normal quality till the highest quality you can get!
Our sending fits every mailbox!
We ship from Monday till Friday!

We ship from Germany and we know how to ship!
It is important for us that all orders arrive in all safety!

SHIPPING TIME
Europe:        	 2 to 7  Business days
Worldwide:       4 to 20 Business days 

REFUND and RESHIP
If orders not arrive please send us a message and we find a solution.
In case of non-arrival we will reship 50\% or a 50\% refund.
Mistakes made in the address-format we will never reship or refund.
New buyers without any order history we never refund or reship.

Please give us some great feedback if you are happy with us!} \\
        \hline
        Price & \thead{17 USD}\\
        \hline
        Shipping from/to & \thead{Germany/Worldwide} \\
        \hline
        Vendor & \thead{\#\#\#\#\#} \\
        \hline
        DWM & \thead{White House} \\
      \end{tabularx}
\label{tab:example_drug_listing_stocks}
\end{table}

\clearpage

\section*{DWMs offering COVID-19 vaccines}

\begin{table}[h]
\footnotesize
\centering
\caption{\textbf{List of DWMs analysed.}}
      \begin{tabular}{lc}
        \hline
Type of products & DWM \\ \hline     
COVID-19 vaccines & \thead{Agartha, Asap, Babylon, Bigblue, \\ Cypher, Dark fox, Hydra, Invictus, Kilos, \\ Liberty, Mgm grand, Recon,  \\ Royal, Televend, The Canadian Headquarters, Torrez, \\ World market, Yakuza, Yukon} \\
COVID-19 related products & \thead{0day.today, Agartha, Asap, Bigblue, \\ Corona, Cypher, Dark fox, Dark0de reborn, \\ Darkmarket, Incognito, Kilos, Liberty, \\ Magbo, Recon, Televend, \\ The canadian headquarters, \\ Torrez, Versus, White house, Yakuza} \\
COVID-19 mentions & \thead{0day.today, 24HoursPPC, ASAP, Agartha, Amigos, \\ Apollon Marketplace, Asean, Atshop, Auction DB, Aurora, \\ Babylon, Big Brother House, BigBlue, Blackhole, CannaHome, \\ Cannabay, Cannazon, Cartel, Cindicator, Connect, \\ Corona, Cypher, Dark Fox, Dark Leak Market, \\ Dark0de Reborn, DarkBay/DBay, \\ DarkMarket, Database, Deep Sea, Deepsy, DutchDrugz, \\ Empire Market, Exchange, FSpros, Faceless, \\ Flugsvamp 3.0, Fullzbuy, Genesis marketplace, \\ HeinekenExpress, Hexablaze, Hookshop, Hydra, Incognito, \\ Invictus, Kilos, Liberty, \\ MEGA Darknet Market, MGM Grand, MagBO, \\ Market Deepmix, Metropolis, Monopoly, \\ Mouse In Box, Namaste LSD, Olux, Opiate Connect, Pentagon, \\ Plati.market, Psylab Seeds, RNJLogs, Recon Search Engine, \\ Royal, Russian Market, SEOclerks, Scans24, \\ Sellix, Shoppy.gg, Silk Road 3.1, Silk Road 4, \\ Tea Horse Road, Televend, The Canadian HeadQuarters, \\ Tor Market, Torrez, UAS, Versus, \\ WTN Market, White House, Willhaben, World Market, \\ Xleet, Yakuza, Yellow Brick marketplace, Yukon} \\
\end{tabular}
\label{market_details}
\end{table}

\clearpage

\section*{Supplementary information}

\begin{table}[!t]
\centering
\caption{\textbf{Vaccine listings detected on DWMs.} Some vendors and DWMs offer vaccines that belong in more than one category.}
      \begin{tabular}{cccc}
        \hline
        \thead{Category} & \thead{Unique istings} & \thead{Vendors} & \thead{DWMs} \\
        \hline
         \thead{\emph{Unspecified vaccines}} & \thead{94}  & \thead{62} & \thead{13} \\ 
        \thead{\emph{Approved vaccines}} &  \thead{75}  & \thead{45} & \thead{7} \\ 
        \thead{\emph{Proofs of vaccination}} & \thead{81} & \thead{43} & \thead{10} \\ 
        \hline
        \thead{Total} & \thead{250}   & \thead{135}  & \thead{19} \\
      \end{tabular}
\vspace{1cm}
\label{Category_Vaccine_listings}
\end{table}

\begin{figure}[h]
  \centering
  \includegraphics[width=15cm]{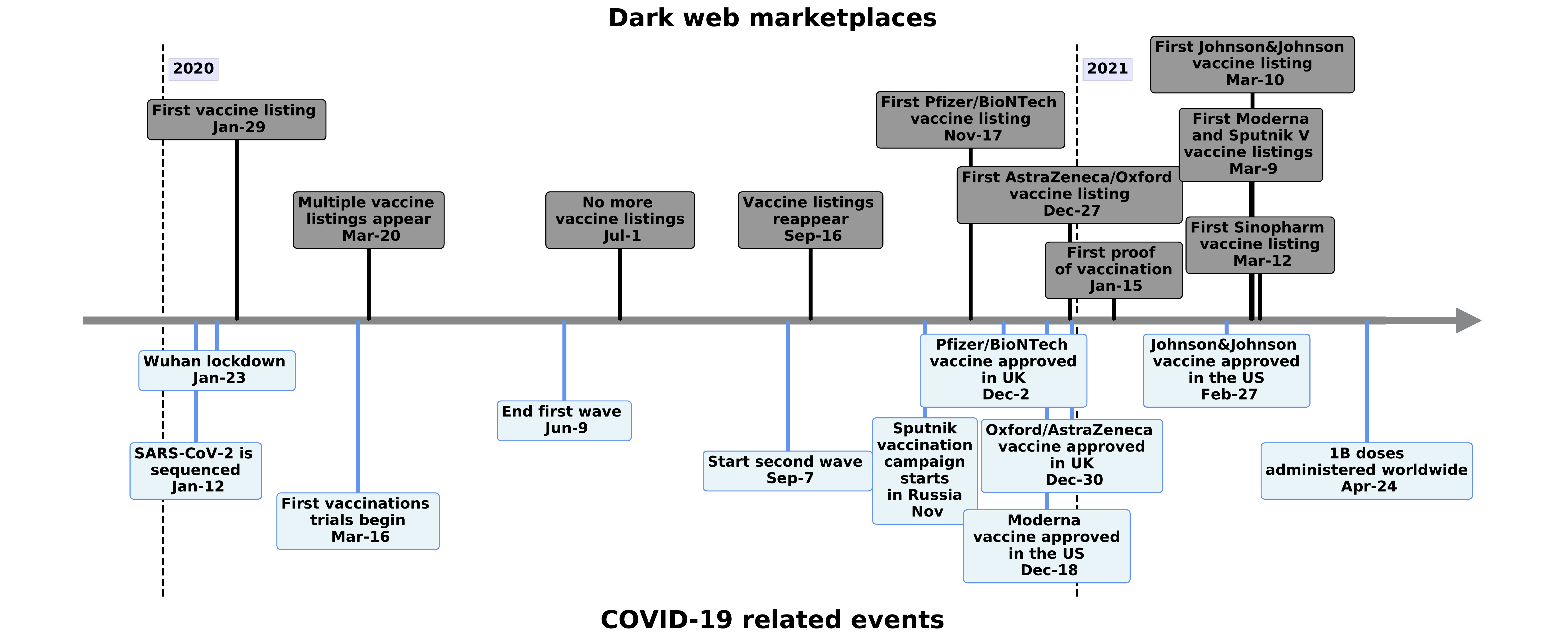}
  \caption{\textbf{Summary of key events related to DWMs and COVID-19.} Availability of listings offering vaccines on dark web marketplaces (top), together with main COVID-19 related events of the vaccination campaign (bottom).}
  \label{timeline_vaccines}
\end{figure}

\begin{figure}[h]
  \centering
  \includegraphics[width=7cm]{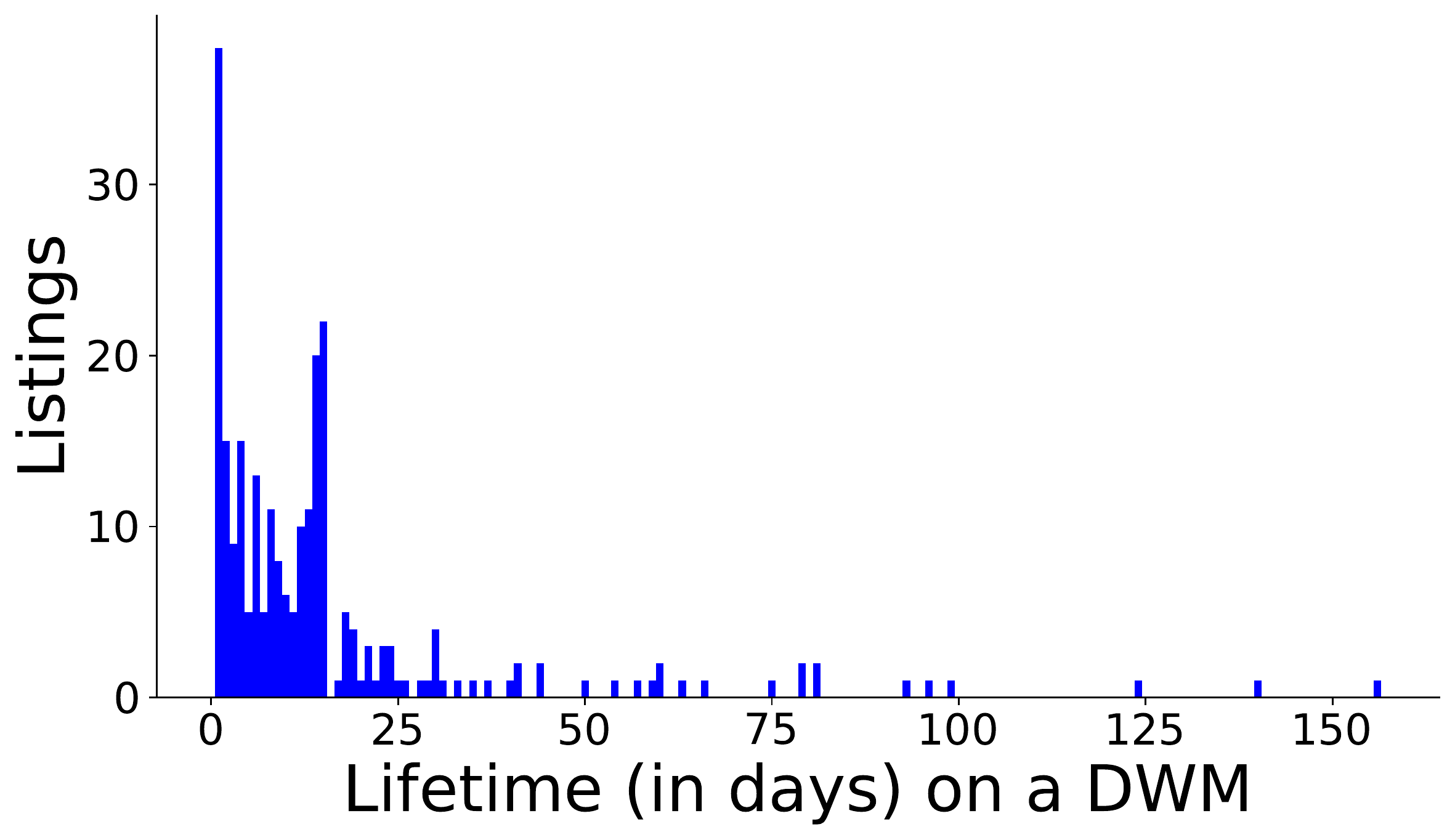}
  \caption{\textbf{Lifetime of listings on a DWM.} Number of days during which listings were active on a DWM.}
  \label{Lifetime_vaccine_listings}
\end{figure}

\begin{figure}[h]
  \centering
  \includegraphics[width=7cm]{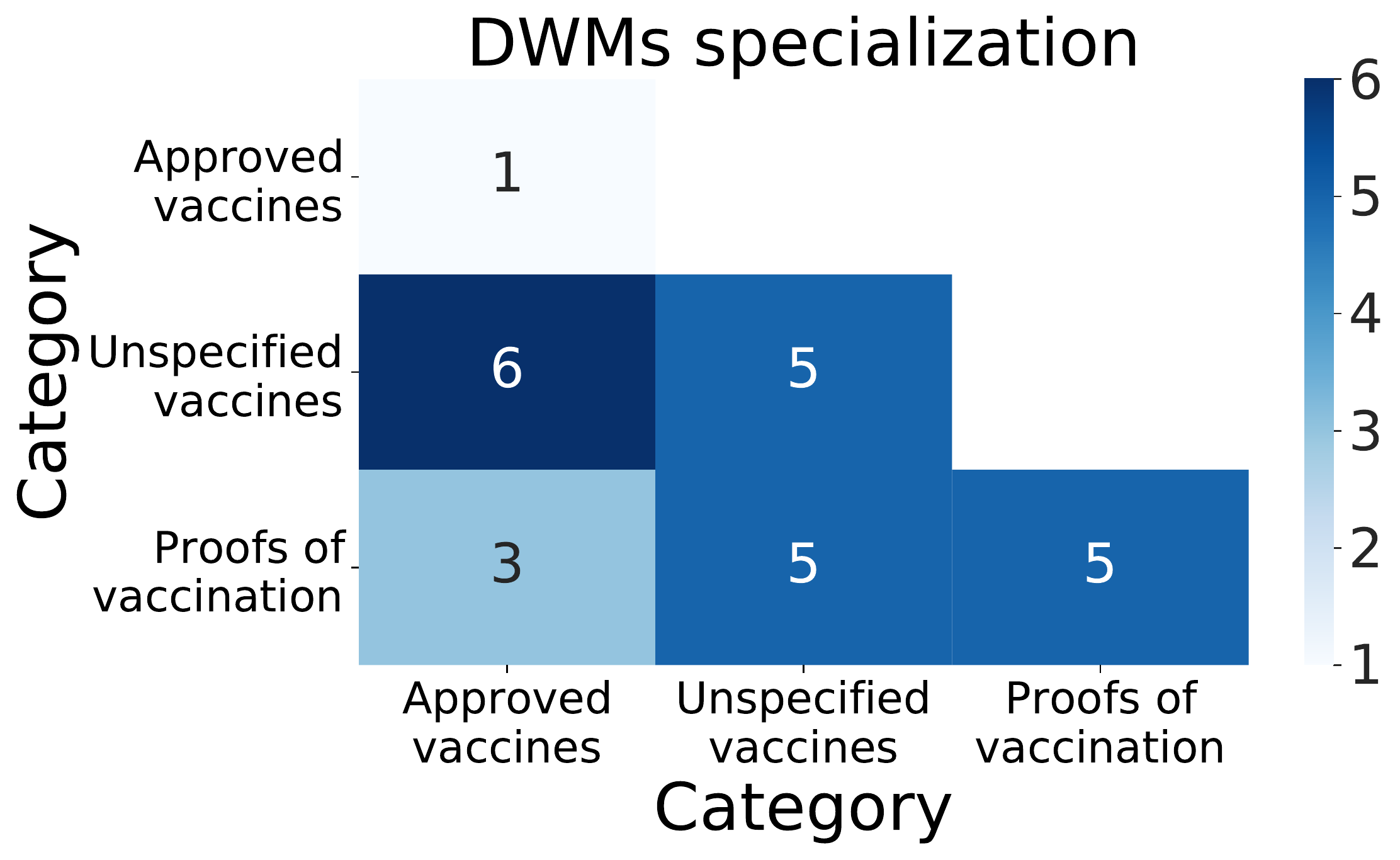}
  \caption{\textbf{Categories of vaccines offered on DWMs.} Number of DWMs offering a vaccine in a given category. Only the lower triangle of the matrix is shown because it is symmetric, where its diagonal represents vendors offering only listings in that category.}
  \label{DWM_listings}
\end{figure}

  \begin{figure}[h]
  \centering
  \includegraphics[width=15cm]{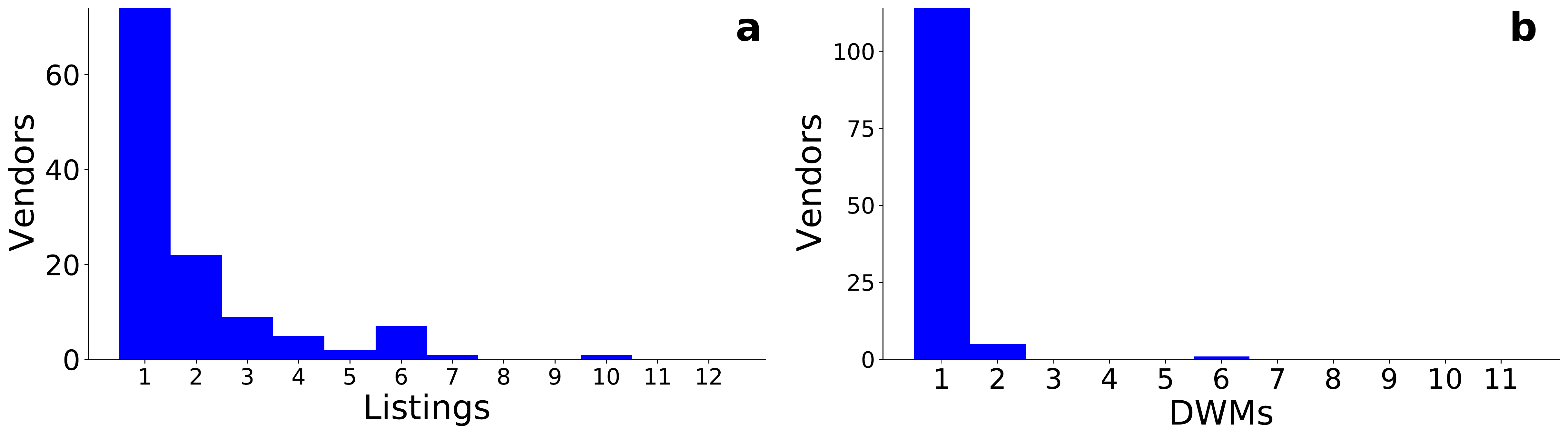}
  \caption{\textbf{Vendor statistics.} Histograms representing the number of vendors offering a certain amount of vaccines listings, in panel (a), and the number of vendors trading in a given amount of DWMs, in panel (b).}
  \label{histograms_vendors}
\end{figure}

\begin{figure}[h]
  \centering
  \includegraphics[width=7cm]{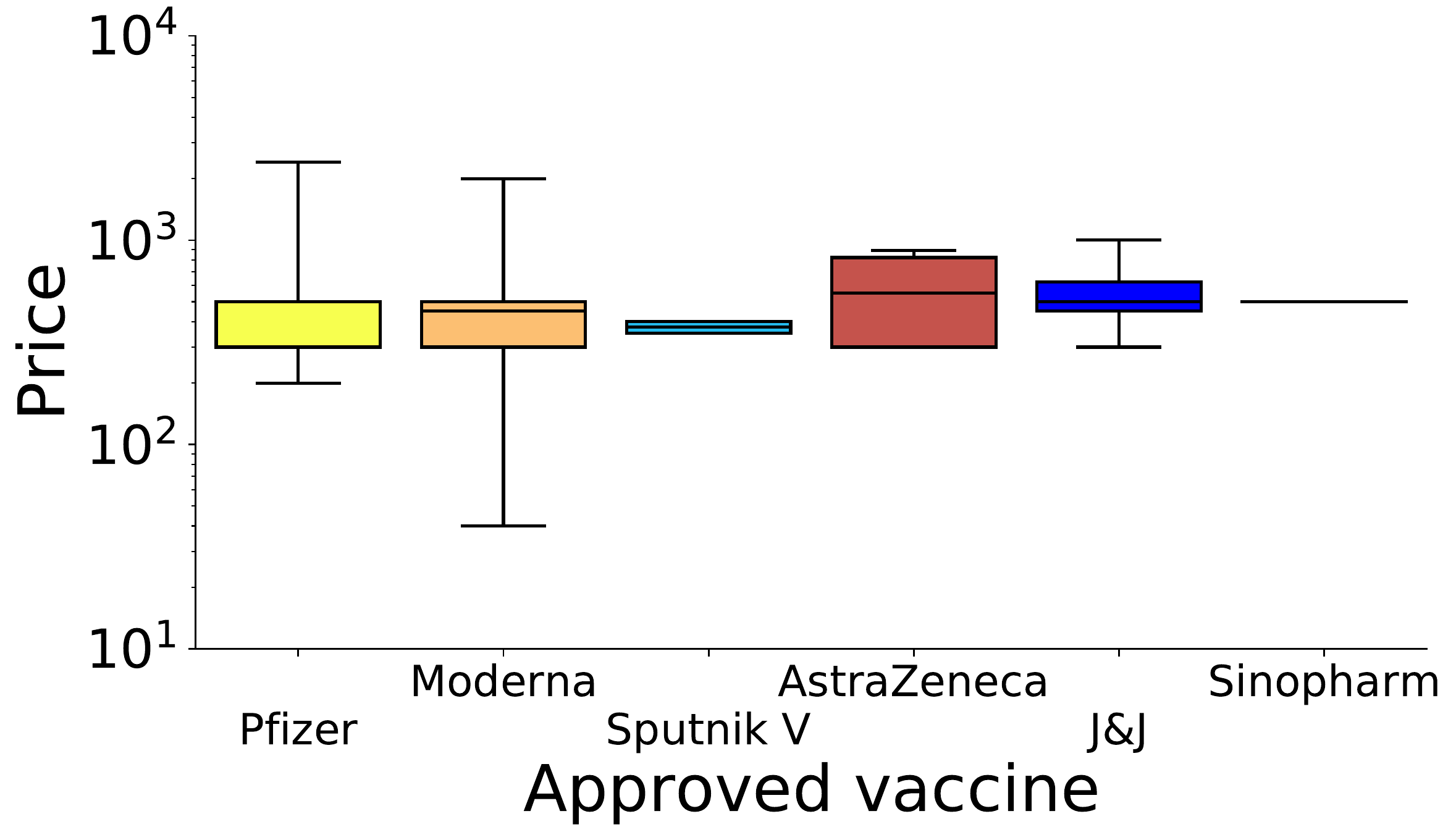}
  \caption{\textbf{Price of COVID-19 \emph{approved vaccines}.} Boxplots of the prices in USD at which vaccines were offered. (a) Price of listings in the three categories considered. (b) Focus on the listings offering approved vaccines. ``J\&J'' stands for Johnson\&Johnson. Horizontal lines represent the median value, box ends the first and third quartiles, and whiskers minimum and maximum values, respectively.}
  \label{Unique_listings_price}
\end{figure}

\begin{figure}[h]
  \centering
  \includegraphics[width=15cm]{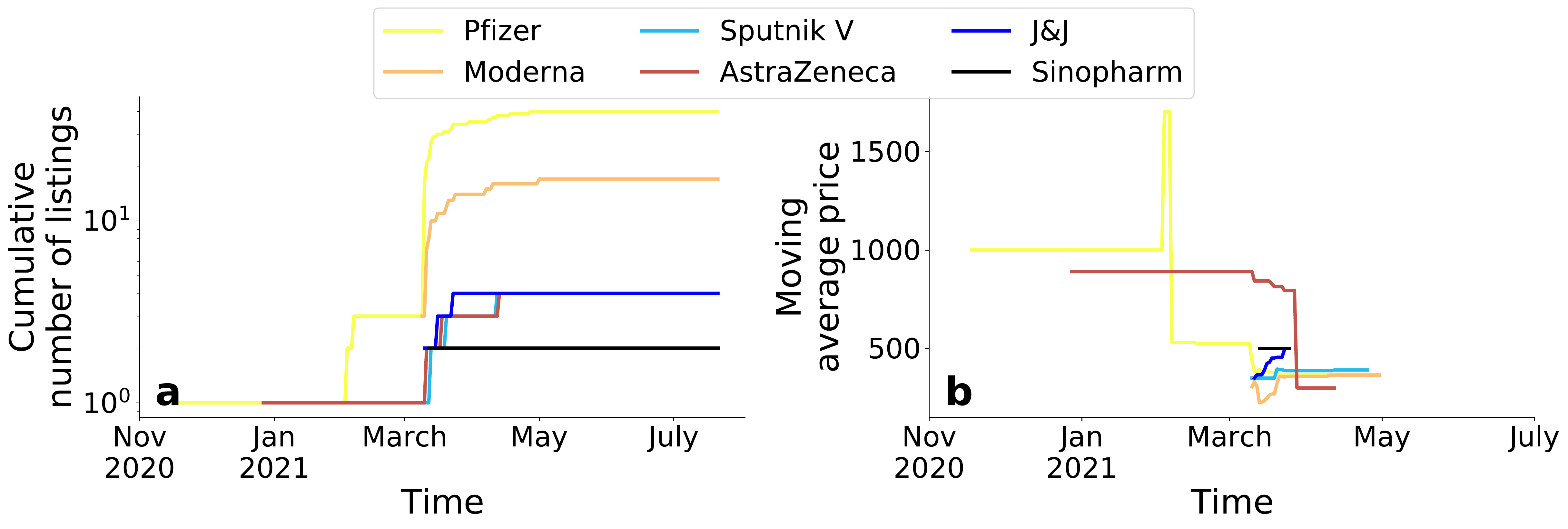}
  \caption{\textbf{Temporal evolution COVID-19 \emph{approved vaccines}.} (a) Cumulative number of listings over time. (b) Average price over time, computed with a 90-days moving window.``J\&J'' stands for Johnson\&Johnson.}
  \label{temporal_evolution_approved_vaccines}
\end{figure}

\begin{table}
\footnotesize
\centering
\caption{\textbf{COVID-19 related products offered on DWMs.} Availability of COVID-19 related products since November 2020.}
\begin{tabular}{lrrrrr}
\hline
Category & Unique listings & Observations &          Median price [USD] &  Vendors & DWMs \\
\hline
Guides on scamming &              62 &         1,046 &          76 &               48 &              16 \\
Malware            &               4 &           19 &             NaN &                3 &               1 \\
Medicines          &              53 &          574 &           45.00 &               30 &              12 \\
PPE                &               8 &           95 &         15 &                3 &               3 \\
Test               &              22 &          130 &          1000 &               11 &               8 \\
Web domain         &              38 &          184 &            4 &               13 &               1 \\
\hline
Total & 187 & 2048 & 50 & 102 & 22
\end{tabular}
\label{tab:covid_19_products}
\end{table}

\begin{figure}[h]
  \centering
  \includegraphics[width=15cm]{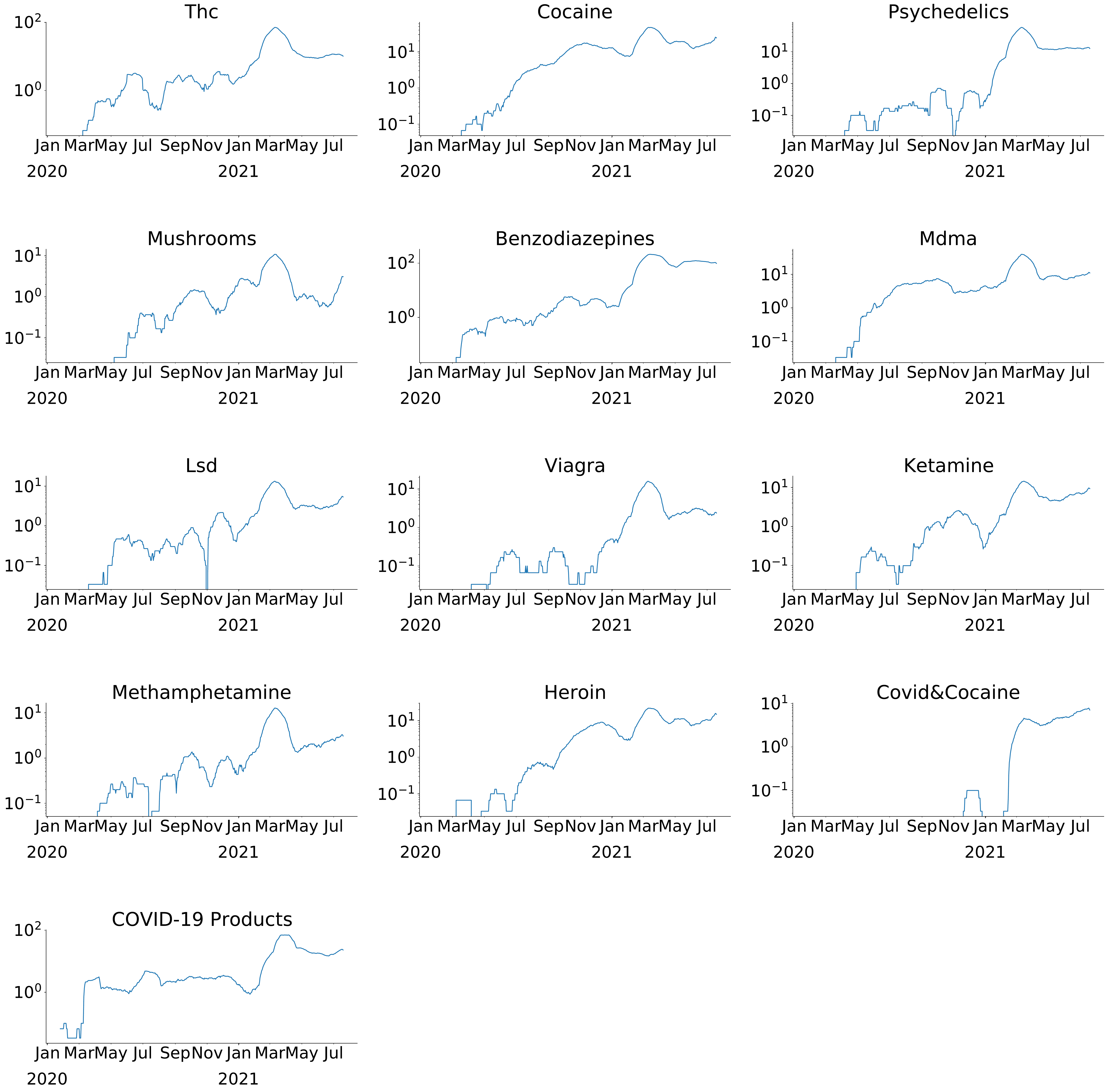}
  \caption{\label{fig:appendix_topic_modelling} \textbf{Time evolution of products mentioning COVID-19.} Number of active listings in time in each product category, according to the clustering described in the main text.}
\end{figure}

\begin{figure}[h]
  \centering
  \includegraphics[width=15cm]{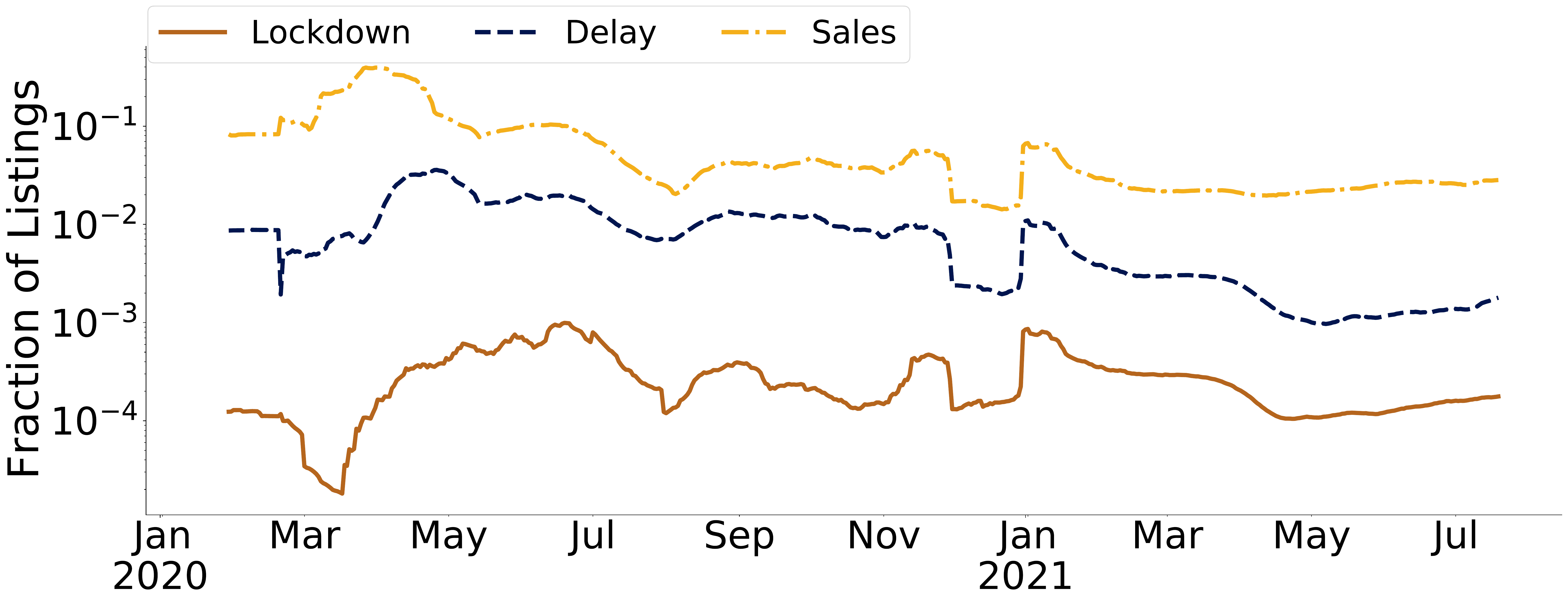}
  \caption{\label{fig:overall_mentions} \textbf{Time evolution of fraction of all listings mentioning COVID-19 related themes.}}
\end{figure}

\end{document}